\documentclass[runningheads]{llncs}
\usepackage{graphicx}
\usepackage[utf8]{inputenc} 
\usepackage[T1]{fontenc}    
\usepackage{hyperref}       
\usepackage{url}            
\usepackage{booktabs}       
\usepackage{amsfonts}       
\usepackage{nicefrac}       
\usepackage{microtype}      
\usepackage{xcolor}         

\usepackage{subfigure}
\usepackage{amsmath}
\usepackage{amssymb}
\usepackage{mathtools}
\usepackage{cleveref}

\begin{document}
\title{Activity Cliff Prediction: Dataset and Benchmark}
%
%
\author{Ziqiao Zhang\inst{1} \and
Bangyi Zhao\inst{1} \and
Ailin Xie\inst{1} \and
Yatao Bian\inst{2}\textsuperscript{$\star$} \and
Shuigeng Zhou\inst{1}\thanks{Corresponding authors.}}
\authorrunning{Z. Zhang et al.}
\institute{Fudan University, Shanghai, China \\
\email{\{zqzhang18, sgzhou\}@fudan.edu.cn}
\and
Tencent AI Lab, Shenzhen, China\\
\email{yataobian@tencent.com}}
\maketitle              
\begin{abstract}
Activity cliffs (ACs), which are generally defined as pairs of structurally similar molecules that are active against the same bio-target but significantly different in the binding potency, are of great importance to drug discovery.
Up to date, the AC prediction problem, i.e., to predict whether a pair of molecules exhibit the  AC relationship, has not yet been fully explored. In this paper, we first introduce ACNet, a large-scale dataset for AC prediction.
ACNet curates over 400K Matched Molecular Pairs (MMPs) against 190 targets, including over 20K MMP-cliffs and 380K non-AC MMPs, and provides five subsets for model development and evaluation. Then, we propose a baseline framework to benchmark the predictive performance of molecular representations encoded by deep neural networks for AC prediction, and 16 models are evaluated in experiments.
Our experimental results show that deep learning models can achieve good performance when the models are trained on tasks with adequate amount of data, while the imbalanced, low-data and out-of-distribution features of the ACNet dataset still make it challenging for deep neural networks to cope with.
In addition, the traditional ECFP method shows a natural advantage on MMP-cliff prediction, and outperforms other deep learning models on most of the data subsets.
To the best of our knowledge, our work constructs the first large-scale dataset for AC prediction, which may stimulate the study of AC prediction models and prompt further breakthroughs in AI-aided drug discovery.
The codes and dataset can be accessed by \href{https://drugai.github.io/ACNet/}{https://drugai.github.io/ACNet/}.

\end{abstract}
%
%
\section{Introduction}
\label{sec:intro}
Current AI-aided drug discovery aims to exploit the rapidly developing deep learning techniques to promote the research \& development of new drugs~\cite{schneider2020rethinking}.
In the process of drug discovery, studies typically start from finding molecules with binding potency against a certain protein target~\cite{dara2021machine,zhong2018artificial}, and these \textit{hit} molecules will be further optimized for desirable bio-activity and ADMET properties in the subsequent phases.
So, molecular property prediction, especially the prediction of bio-activity, is  one of the fundamental tasks in AI-aided drug discovery~\cite{zhang2022gans}.

Recently, many molecular property prediction models have been proposed in the literature~\cite{coley2017convolutional,jaeger2018mol2vec,peng2020top,song2020communicative,xiong2019pushing,ying2021transformers,zhang2021fragat} and have shown good performance on some benchmarks~\cite{hu2021ogb,wu2018moleculenet}.
However, the behaviors of these models cannot always meet the requirements of pharmacists when they are applied to predicting the binding affinities against certain targets in practice~\cite{kim2021comprehensive}.
Performance degradation occurs for a variety of reasons, and Activity Cliffs (ACs) are an important issue ~\cite{van2022exposing}.

Conceptually, ACs are defined as pairs of structurally similar compounds that are active against the same bio-target but significantly different in binding potency~\cite{dimova2016advances}.
This phenomenon indicates that a trivial chemical modification may lead to dramatic change of bio-activity~\cite{stumpfe2019evolving}, and its impact on drug discovery is two-sided. 
On the one hand, ACs can serve as a valuable source of information for pharmacists and chemists to model and analyze Quantitative Structure-Activity Relationship (QSAR) and an important knowledge repository for understanding the binding features of protein binding pockets~\cite{leung2012methyl,mousa2022exploiting,stumpfe2019evolving} and studying molecular similarity relationships~\cite{dimova2016advances}.
In addition, taking AC phenomenon as the starting point of drug discovery studies is more likely to find candidates of higher binding affinity~\cite{dimova2013medicinal,stumpfe2013compound} and greatly boosts the efficiency for new drug discovery~\cite{abad2005ligand,dimova2016advances,leeson2007influence}.
Therefore, ACs play an important role in accelerating the exploration of active molecules in the early stage of drug discovery and design.

On the other hand, at the later stages of lead optimization phase, the AC phenomenon is undesirable since it may lead to significant loss in bio-activity when optimizing multiple ADMET properties by chemical modifications~\cite{stumpfe2019evolving}.
Furthermore, the occurrence of ACs acts as an exception to the fundamental hypothesis underlying molecular property prediction models that \textit{similar compounds are likely to have similar properties}~\cite{van2022exposing}.
So, the molecules involved in AC relationships would be hard cases for deep learning models to cope with~\cite{stumpfe2019evolving}.

Considering that ACs play an important role in drug discovery~\cite{stumpfe2019evolving}, it is necessary to develop methods for predicting whether a pair of compounds would exhibit AC relationship, which is the task of AC prediction.
Although the AC issue has been studied for decades in the Computer-Aided Drug Discovery (CADD) community, AC prediction by ML/DL methods has not been fully explored in the literature. Up to now, only a few works have focused on this task~\cite{heikamp2012prediction,horvath2016prediction,iqbal2021prediction,de2014prediction}.
What is worse, there is no benchmark dataset dedicated to this task.
As ImageNet~\cite{deng2009imagenet} has demonstrated, in the deep learning era, benchmark datasets can serve as more than a simple collection of data, but a critical trigger for the community to develop better solutions and promote technical breakthroughs~\cite{wu2018moleculenet}.
So, to stimulate the study of AC prediction, a benchmark dataset is urgently required. Furthermore, 
as deep representation learning models have achieved good performance on molecular property prediction, it is worthy of exploring how well these models can perform on AC prediction.

In this work, following the footsteps of ImageNet~\cite{deng2009imagenet} and MoleculeNet~\cite{wu2018moleculenet}, we first construct a dataset named ACNet for the development and evaluation of deep learning models designed for AC prediction.
ACNet curates over 400K Matched Molecular Pairs (MMP) against 190 targets, and provides five subsets with imbalanced, low-data and out-of-distribution features. Then, we develop
a baseline framework for evaluating the predictive performance of molecular representations encoded by deep learning models for AC prediction.
16 models are evaluated in extensive experiments, which benchmark the performance of these models and reveal the challenge and necessity of the ACNet dataset.

\section{Related Work}
\label{sec:related_work}

Heikamp~\textit{et al.}~\cite{heikamp2012prediction} proposed the first approach to predict AC relationships between molecules.
Support Vector Machine (SVM) is used for MMP-cliff prediction.
For each MMP, molecules are firstly divided into one core substructure and two transformation substructures (i.e., the substituents), and fingerprints are used for encoding these substructures into representation vectors.
Three carefully tailored compound pair-based kernel functions are designed to capture the similarity between two MMPs.
The proposed model is exploited to predict AC relationships on compounds against 9 different targets.
Leon~\textit{et al.}~\cite{de2014prediction} tried to predict the potency changes of MMPs, which can be considered as a regressive AC prediction task.
Similar to previous work, molecules in an MMP are firstly split into three substructures and encoded separately.
Support Vector Regression is used for prediction.

Later, Horvath \textit{et al.}~\cite{horvath2016prediction} introduced the Condensed Graph of Representation (CGR) method for AC prediction.
CGR is originally designed for modeling chemical reactions, which uses a single molecular graph to synthetically characterize a chemical transformation including both conventional bonds and ``dynamic'' bonds.
Descriptors of a pair of molecules in an MMP are concatenated with a specialized designed method to distinguish the core and substituents.

Recently, Iqbal \textit{et al.}~\cite{iqbal2021prediction} introduced deep models to AC prediction.
Molecules of an MMP are split first.
Then, the three substructures are transformed into three images by the RDKit toolkit and are concatenated into one image.
Convolutional Neural Networks (CNN) are used as encoders to embed the image into latent vectors and make predictions of AC relationship.
Data against three targets are used in the experiments, and the proposed CNN-based model can achieve good performance on these tasks, with AUC-ROC up to 0.97.

All these works above have not addressed the concerns mentioned in Sec.~\ref{sec:intro}.
First, the datasets used in these works are self-collected and not public available, so a benchmark dataset for AC prediction is still lacking.
Second, all of these works exploit a core/substituent splitting method, 
predictions are made by representations of the substructures, rather than the intact molecules.
So, they cannot answer the question about the predictive performance of deep molecular representations on AC prediction.
Instead, this paper tries to cope with these concerns not addressed in the literature.

\section{The ACNet Dataset}
\label{sec:ACNet}
In this section, to promote the study on AC prediction models, a large-scale dataset for AC prediction is built.

\subsection{Activity Cliff Definition}
Activity Cliffs are generally defined as pairs of structurally similar molecules that are active against the same bio-target but have large discrepancy in binding potency~\cite{stumpfe2019evolving}, but there is still no consensus about the molecular similarity criteria.
Researchers have proposed different criteria for identifying ACs, including Fingerprint cliffs~\cite{stumpfe2012exploring}, Chirality cliffs~\cite{schneider2018chiral}, Scaffold/R-group cliffs~\cite{Hu2012extending}, MMP cliffs~\cite{hu2012mmp}, Isomer/MMP cliffs~\cite{Hu2020Introducing}, Analog Pair cliffs~\cite{stumpfe2019introducing}, 3D cliffs~\cite{hu2015extension}, etc.
MMP is a pair of compounds that can be only distinguished by a chemical modification at a single site, which is well consistent with the concept of \textit{similar} molecules intuitively, therefore the MMP cliffs have received more attention in the literature~\cite{heikamp2012prediction,iqbal2021prediction}.
Consequently, in this work, Matched Molecular Pair is selected as the similarity criterion of Activity Cliffs.

\subsection{Data Collection}
\begin{figure}[htbp]
\begin{center}
\centerline{\includegraphics[width = \linewidth]{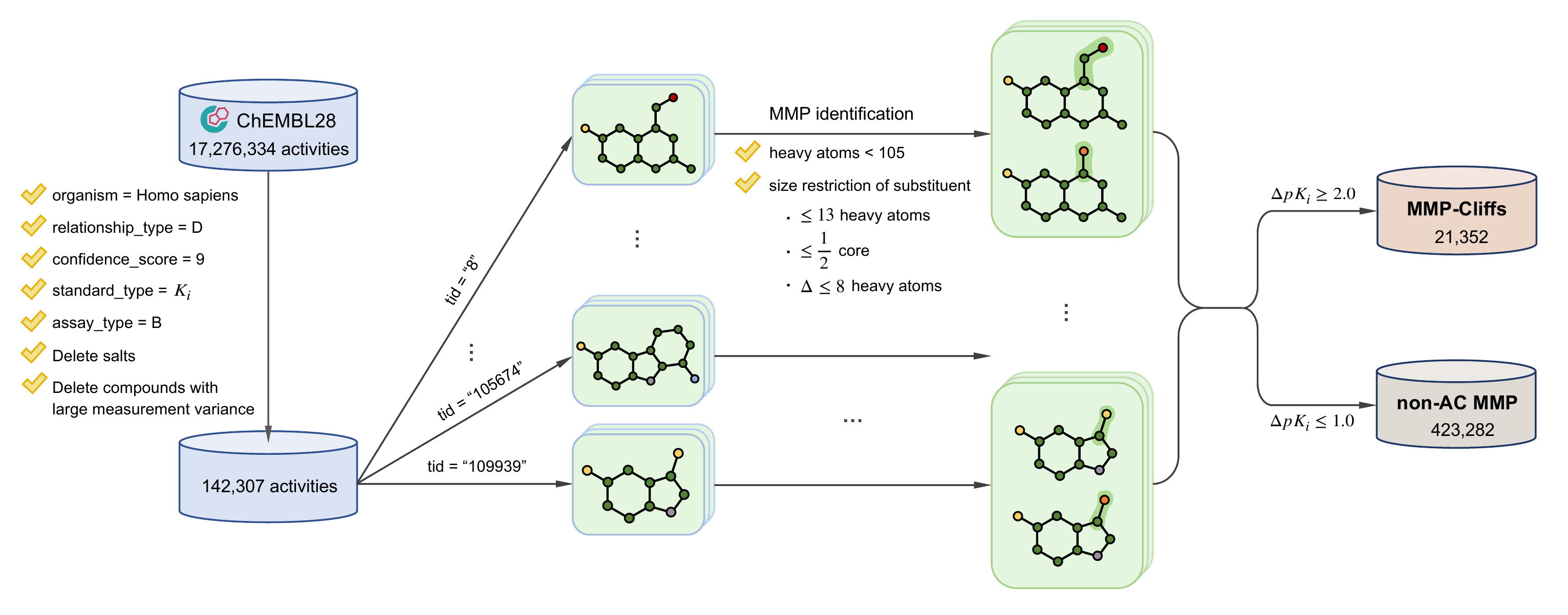}}
\caption{The flowchart of data collection of the ACNet dataset. Activities are gathered from ChEMBL28 database and screened by the proposed rules. MMPs against each target are identified by the algorithm proposed by Hussain~\textit{et al.}~\cite{hussain2010computationally} with size restrictions of substituents.}
\label{fig:ACNet}
\end{center}
\end{figure}

The data in ACNet are collected from publicly available database ChEMBL (version 28)~\cite{gaulton2017chembl}.
Over 17 millions of activities, each of which records the binding affinity of a molecule against a target, are screened by the rules shown in Fig.~\ref{fig:ACNet}.
Compounds trialed against single human targets (organism = Homo sapiens) in direct interaction binding assays (assay\_type = B, relationship\_type = D) at the highest assay confidence (confidence\_score = 9) are reserved to construct the dataset.
Assay-independent equilibrium constants ($pK_i$ values) are used as the potency measurement.
Salt compounds are discarded.
Multiple measurements of the same compound against the same target are averaged if all values fall within the same order of magnitude.
Otherwise, this activity record is discarded.
The reason to adopting these screening rules are explained in Supplementary~\ref{sec:rules}.
As a result, 142,307 activities are included in our dataset.

Next, to identify pairs of molecules exhibiting AC relationships against each target, all of the activities are treated separately according to the \textit{tid}.
All possible MMPs are identified by the algorithm proposed by Hussain~\textit{et al.}~\cite{hussain2010computationally}.
Size restrictions of substituents are also applied.
First, a substituent is restricted to contain at most 13 heavy atoms, and the core has to be at least twice as large as the substituent.
Second, the difference between the substituents of an MMP is restricted to be at most 8 heavy atoms.
These restrictions make the identified MMPs consistent with the typical structural analogues in practice~\cite{iqbal2021prediction} (see Supplementary~\ref{sec:rules}).

For each MMP, if the difference in potency is greater than 100-fold (i.e., $\Delta pK_i \geq 2.0$), then the MMP is considered as an MMP-cliff with a positive label.
When the potency difference is lower than 10-fold, then the MMP is denoted as a non-AC MMP with a negative label.
This criterion involves a distinct margin between the potency differences of positive samples and those of negative samples, so that the influence of the observational error induced by the source assays can be limited.

Based on the above-mentioned data collection and screening rules, we eventually obtain a total of 21,352 MMPs exhibiting AC relationships, and 423,282 negative non-AC MMPs.
Examples of the data in ACNet are shown in Fig.~\ref{fig:dataset}.


\subsection{Data Organization}
In ACNet, each sample represents an MMP, and the label of each sample indicates whether it exhibits an AC relationship against a certain target.
It is intuitive to organize the samples against different targets into different prediction tasks.
To construct dataset for each task, positive samples and negative samples against the same target should be gathered first.
In this step, a threshold is applied to screening out the tasks with extremely few positive samples, since the scarceness of positive samples brings little information of the tasks, and it is too tough for a deep learning model to be trained on these tasks.
We set this threshold to be 10.


Under this configuration, ACNet contains MMPs against 190 targets, i.e., 190 tasks.
And the numbers of samples in each task range from 36 to 26,376.
As the number of tasks is large and the data volume of each task varies greatly, for the convenience of model evaluation and comparison, we divide the original 190 tasks into several groups according to the task size.
By default, tasks with more than 20,000 samples are organized as the Large subsets, tasks with 1,000 to 20,000 samples forms the Medium subsets, and tasks with 100 to 1,000 samples are curated as the Small subsets, finally tasks with less than 100 samples constitute the Few subsets.

A summary of the data organization of our dataset is given in Tab.~\ref{tab:arrangement}.
And more statistical information of the dataset is introduced in Appendix~\ref{sec:moreinformation}.
From the figures in Appendix~\ref{sec:moreinformation}, we can see that ACNet shows imbalanced and low-data features.
For convenience, we refer to the Large, Medium, Small subsets collectively as \textit{ordinary subsets} in the following.

\begin{table}[htbp]
  \caption{Statistics of different data subsets in ACNet.}
  \label{tab:arrangement}
  \centering
  \begin{tabular}{cccc}
    \hline
    Subsets & \#tasks & threshold & \#samples\\
    \hline
    Large & 3 & $>20000$ & 72,233\\
    Medium & 64 & $[1000,20000]$ & 275,927\\
    Small & 110 & $[100,1000]$ & 53,084\\
    Few & 13 & $<100$ & 835\\
    \hline
    Mix & 1 & - & 278,367\\
    \hline
  \end{tabular}
\end{table}

\subsection{Domain Generalization via Target Splitting}


In the previous subsection, samples against different targets are organized into different predictive tasks.
Models can be trained on these tasks separately to learn knowledge about the chemical modifications leading to large potency difference against a certain target.
However, there may be common knowledge unveiling what chemical modifications are more probable to cause a large difference in binding potency no matter the targets, and an AC prediction model may be expected to learn such common knowledge to better understand the latent principles behind the AC phenomenon and structural similarities between molecules from a potency-based perspective.


Motivated by this, we propose an extra Mix subset where all of the samples against different targets are organized into a single task to construct a \textit{mixed} dataset.
To avoid ambiguity, conflicting MMPs, i.e. MMPs showing different AC relationships on tasks of targets with known labels, are discarded.
And for repeated MMPs, i.e. MMPs showing the same AC relationships on tasks with known labels, only one sample of each MMP is remained.
The number of samples in the Mix subset is 278,367, as shown in Tab.~\ref{tab:arrangement}.
To force deep models to learn common knowledge from the Mix subset, a \textbf{target splitting} method is proposed, following the idea of the scaffold splitting method in the molecular property prediction tasks.
Specifically, when splitting the Mix subset into train/valid/test sets, samples that against the same target must be split into the same set.

The Mix subset consists of samples against different targets, of which the data distribution are discrepant. 
And the target splitting method makes the data for training and evaluating be sampled from different data distributions.
In this case, the prediction task of the Mix subset is a \textbf{domain generalization} problem, which consequently brings out-of-distribution (OOD) feature to ACNet.

Domain generalization, i.e. out-of-distribution generalization, focuses on the problem that learning a model from one or several different but related domains (data distributions) to generalize on unseen testing domains, which is ubiquitous in real-world scenarios~\cite{wang2022generalizing}.
As traditional deep learning models are trained based on the independent identically distributed (i.i.d.) hypothesis, i.e., data for training and testing are sampled independently from identical distribution~\cite{zhang2021deep}, tasks with OOD feature is of great challenge for deep learning models and will lead to performance degradation in \textit{distribution shifting} situations~\cite{2022arXiv220109637J}.
So, 
the OOD feature of the Mix subset will dramatically increase the difficulty for deep models.

Notably, for the tasks of the subsets except Mix, since the scaffolds of the two molecules of an MMP are mostly different, we can use only the random splitting method instead of the scaffold splitting.

\section{A Baseline Framework for AC Prediction}
\label{sec:baseline_framework}
As introduced in Sec.~\ref{sec:related_work}, previous efforts on predicting Activity Cliffs exploit a feature engineering pre-processing method that the molecules in an MMP are split into one core and two substituents, and these three parts are encoded separately.
However, this splitting approach can only be adopted to cope with MMP-cliffs, where only one single site's substituents are different.
As the criteria of molecular similarity in AC definitions are various, this splitting approach will not work when other criteria are adopted, such as Tanimoto similarity criterion or 3D similarity criterion.
In addition, predictions of existing works are made by representations of substructures, so that the prediction accuracy cannot reveal the real predictive performance of the whole molecular representations, where the information of intact molecules are encoded.
So, to answer the question raised in Sec.~\ref{sec:intro}, in this section we propose a baseline framework as benchmark for AC prediction.

\begin{figure}[tbp]
\begin{center}
\centerline{\includegraphics[width = \linewidth]{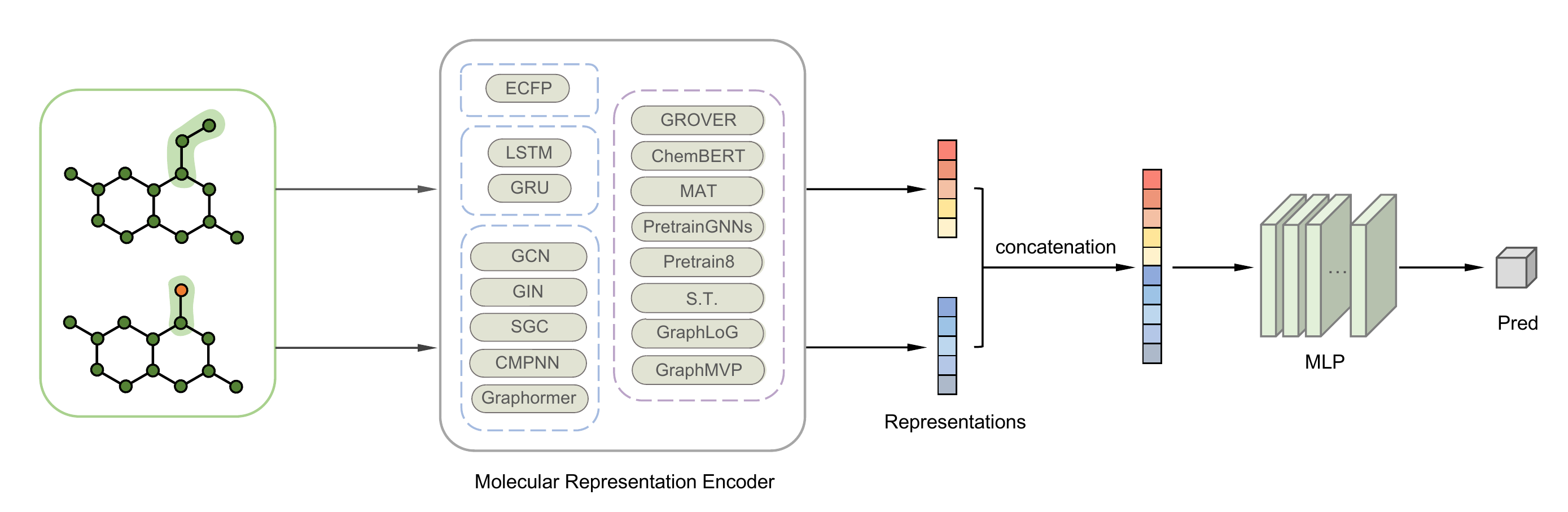}}
\caption{A baseline framework for AC prediction. Various existing deep learning models are used as backbone encoders to extract molecular representations.}
\label{fig:framework}
\end{center}
\end{figure}

As shown in Fig.~\ref{fig:framework}, a backbone deep learning model is leveraged as an encoder to extract molecular representations of the two intact compounds in an MMP, and these representations are concatenated as input of an MLP prediction head to predict the AC relationship of these compounds.
This baseline framework predicts AC relationship based on the representations of intact molecules, which conforms to the concern of this work. 
In addition, this baseline framework makes no \textit{a priori} assumption about the similarity criteria, so that it is compatible to any definition of Activity Cliffs.
In this work, the predictive performance of 16 backbone models are evaluated in experiments. Details of these experiments are given in Tab.~\ref{tab:models}. Note that this framework is open to include any other deep representation models. 


\section{Experimental Evaluation}
\label{sec:exp}
ACNet enables researchers to develop and evaluate deep learning models for the AC prediction task.
In this section, we train models under the baseline framework on the ACNet dataset to evaluate how existing backbone deep models on the AC prediction task.
As the existing works introduced in Sec.~\ref{sec:related_work} cannot meet the goal of this study,
they are not involved in the comparison of this benchmark.
However, experiments in Supplementary~\ref{sec:cs_splitting} reveal that the core/substituent splitting method can lead to prominent accuracy improvement.
So, the splitting method can serve as an add-on skill in practice.

\subsection{Results on the Ordinary Subsets}
Experiments are first conducted on the \textit{ordinary subsets}, i.e., Small, Medium and Large subsets, which cover most of the tasks in the ACNet dataset.
Models are trained with the same set of hyperparameters for each task of the subsets, and the random splitting ratio is 8:1:1.
AUC-ROC is used as the measurement for prediction accuracy.
The average accuracy of all tasks of a subset is considered as the performance of the model on this subset.
Experiments are repeated three times and the average performance is reported with error bar.
The results are shown in Tab.~\ref{tab:baseline_exp}, and the best score of each subset is in bold.

From Tab.~\ref{tab:baseline_exp}, we can see that if the models are trained on tasks with adequate amount of data, e.g. tasks in Large and Medium subsets, most of these baseline models can achieve good performance.
And the simplest ECFP+MLP model can achieve an outstanding 0.984 AUC on the Large subset and 0.910 AUC on the Medium, which outperforms all of the other complex deep models.
The reason behind the eye-catching performance of ECFP+MLP can be explained by the natural advantage of ECFP in embedding similar molecules of MMPs.

The function that an MMP-cliff prediction model tries to fit is the relationship between the chemical modification of two substituents and the potency difference, with the core structure as a context.
When using the intact-molecular representations for prediction, the context core part and the substituent part are embedded integrally in the representation vector, therefore the prediction head have to implicitly learn to identify these parts.
However, in ECFP, each bit represents the occurrence of a certain circular substructure in the molecule~\cite{rogers2010extended}.
So, for two molecules in an MMP, the constant core and variant substituent are explicitly marked by the bits in ECFP.
This feature of ECFP will enable the downstream prediction head to explicitly identify the constant part and the modified part of two similar molecules, therefore achieve better performance.

As the number of samples of the tasks in the Small subset is not large enough, it leads to an obvious degradation of prediction accuracy of the baseline models.
The representation encoders implemented by SMILES-based and graph-based models need to be trained from scratch, so that the low-data feature of Small subset has greater impact to the performance of these models than that of the ECFP+MLP.
In addition, due to the small amount of data and the substantially imbalanced data distribution, over-fitting of these models can be observed, leading to large difference between the accuracies on train/valid/test sets of each task.
So, the Small subsets are more challenging.




\begin{table}[htbp]
  \caption{Performance of baseline models on the ordinary subsets.}
  \label{tab:baseline_exp}
  \centering
  \begin{tabular}{cccc}
    \hline
    Model     &  Large  & Medium & Small \\
    \hline
    ECFP+MLP & \textbf{0.984 $\pm$ 0.003} & \textbf{0.910 $\pm$ 0.005} & \textbf{0.890 $\pm$ 0.004}\\
    LSTM & 0.964 $\pm$ 0.006 & 0.855 $\pm$ 0.012 & 0.812 $\pm$ 0.008\\
    GRU & 0.975 $\pm$ 0.001 &  0.846 $\pm$ 0.004 & 0.803 $\pm$ 0.007\\
    GCN & 0.977 $\pm$ 0.001 & 0.862 $\pm$ 0.005 &  0.852 $\pm$ 0.003\\
    GIN & 0.723 $\pm$ 0.004 & 0.607 $\pm$ 0.001 & 0.625 $\pm$ 0.003 \\
    SGC & 0.979 $\pm$ 0.001 & 0.858 $\pm$ 0.001 & 0.835 $\pm$ 0.012 \\
    CMPNN & 0.983 $\pm$ 0.001 & 0.869 $\pm$ 0.005 & 0.821 $\pm$ 0.013\\
    Graphormer & 0.971 $\pm$ 0.006 & 0.822 $\pm$ 0.012 & 0.753 $\pm$ 0.013\\
    \hline
  \end{tabular}
\end{table}

\subsection{Results on the ``Few'' Subset}

Due to the limited data of the Few subsets, it is impossible to train deep learning models to extract representations of compounds from scratch.
Following the pretrain-finetune paradigm in few-shot learning, self-supervised pre-trained models (PTMs), which are trained on large-scale unlabeled data by carefully designed pretext tasks to extract representations of molecules, are exploited here.
Specifically, well-trained PTMs are fixed as encoders to extract molecular representations, and only the down-stream MLPs are finetuned on the training set.
The performance results of 8 PTMs and ECFP are given in Tab.~\ref{tab:pretrain_exp}.
As the performance are overlapped due to the relatively large standard deviation, the top-2 scores are in bold.


The results in Tab.~\ref{tab:pretrain_exp} further reflect the advantage of ECFP as molecular representations in MMP-cliff prediction, since it achieves the second best prediction accuracy on the Few subset.
On the contrary, although the state-of-the-art PTMs have been trained by large amount of unlabeled data, performance of most PTMs is even worse than the ECFP.
These findings further demonstrate the difficulty of the Few subset in the proposed ACNet dataset.

\begin{table}[htbp]
  \caption{Performance results of PTMs and ECFP on the Few subset.}
  \label{tab:pretrain_exp}
  \centering
  \begin{tabular}{cccc}
    \hline
    Models & AUC & Models & AUC\\
    \hline
    GROVER \cite{rong2020self} & 0.753 $\pm$ 0.010 & ChemBERT \cite{kim2021merged} & 0.656 $\pm$ 0.029 \\
    MAT \cite{maziarka2020molecule} & 0.730 $\pm$ 0.069 & Pretrain8 \cite{chen2021extracting} & 0.782 $\pm$ 0.031 \\
    PretrainGNNs \cite{hu2019strategies} & 0.758 $\pm$ 0.015 & S.T. \cite{honda2019smiles} & \textbf{0.822 $\pm$ 0.022} \\
    GraphLoG \cite{xu2021self} & 0.751 $\pm$ 0.040 & GraphMVP \cite{liu2021pre} & 0.724 $\pm$ 0.026 \\
    ECFP & \textbf{0.813 $\pm$ 0.024}  \\
    \hline
  \end{tabular}
\end{table}

\subsection{Results on the ``Mix'' Subset}
The results of the experiments on the Mix subset with random splitting and target splitting methods are presented in Tab.~\ref{tab:mix_exp}, and the best scores are in bold.

We can see that the degradation of prediction accuracy, which up to over 40 percentage, is significant when using the target splitting method.
The ECFP+MLP, which is eye-catching in previous experiments, fails at this time, since the advantage of ECFP cannot afford the weakness of the generalization ability of a simple MLP under the OOD context.
And although the GCN model achieves the best performance on this domain generalization task, the 0.579 AUC indicates that we cannot assume that this model has learnt the common latent mechanism behind the ACs phenomenon.
Moreover, even the SOTA model Graphormer cannot exhibit good generalization ability when coping with this domain generalization task.
These findings show that the Mix subset with target splitting is of great challenge to deep learning models.
Algorithms designed for OOD issues~\cite{2022arXiv220109637J} are required to solve this problem.


\begin{table}[htbp]
  \caption{Performance of baseline models on the Mix subset.}
  \label{tab:mix_exp}
  \centering
  \begin{tabular}{ccc}
    \hline
    Model     &  Random splitting  &  Target splitting \\
    \hline
    ECFP+MLP & 0.960 $\pm$ 0.001 & 0.522 $\pm$ 0.015 \\
    GRU & \textbf{0.963 $\pm$ 0.001} & 0.563 $\pm$ 0.011\\
    GCN & 0.941 $\pm$ 0.001 & \textbf{0.579 $\pm$ 0.031}\\
    Graphormer & 0.935 $\pm$ 0.025 & 0.520 $\pm$ 0.014\\
    \hline
  \end{tabular}
\end{table}



\section{Discussions and Future Work}
\label{sec:discussion}
In this paper, we address the Activity Cliff prediction issue, which is important in drug discovery, but has not yet been rigorously studied. 
A large-scale benchmark dataset is first built for the AC prediction task.
The ACNet dataset consists of over 400K MMPs against 190 targets, and provides five subsets for model development and evaluation.
Then, a baseline framework for AC prediction is proposed, with which 16 baseline models are evaluated on ACNet.
The experimental results reveal the outstanding performance of ECFP over other deep models on MMP-cliffs prediction.
Moreover, the imbalanced, low-data and OOD features of the ACNet dataset make it of great challenge for deep learning models.

As for future work, ACNet will be persistently updated to gather more up-to-date activity data from the latest version of ChEMBL and provide more choices of configurations for customized dataset organization, e.g. dividing tasks by the category of targets. And
other molecular similarity criteria will be implemented to identify more categories of ACs.

\clearpage
\appendix

\begin{center}
	\LARGE \bf {Supplementary Materials of ACNet}
\end{center}

\setcounter{table}{0}
\renewcommand{\thetable}{S\arabic{table}}
\setcounter{figure}{0}
\renewcommand{\thefigure}{S\arabic{figure}}
\setcounter{equation}{0}
\renewcommand{\theequation}{S\arabic{equation}}


\section{Explanation of data screening rules}
\label{sec:rules}
The screening rules used in the data collection process mainly focus on three goals: data uniformity, data correctness and data consistency with reality.

First, \textit{relationship\_type = D} indicates direct protein target assigned, rather than homologous protein target. 
This rule will guarantee the correctness and consistency with reality of the data.
Also, DNA and non-molecular targets will be discarded in this phase to guarantee the uniformity of data.
The \textit{Confidence\_score = 9} indicates the highest assay confidence in the ChEMBL database, which is used to guarantee the correctness of the data.
\textit{Organism = Homo sapiens} indicates human targets, which are the targets that drugs are developed to bind with in practice.
In addition, only compounds with $K_i$ measurement available were taken into account, since the value of $K_i$ is independent to the enzyme concentration and substrate concentration used in the experiment.
It makes the $K_i$ value accurate and appropriate to be compared, unlike other measurements, e.g. $IC_{50}$.
Finally, \textit{assay\_type = B} means that this is a binding assay other than an ADME, functional or toxicity assay.
This rule is for the uniformity of the data since most of the assays remained after the filtering will fell into this type.
For the same reason, salts are also deleted.

In the process of identifying MMP-cliff, we remain compounds with less than 105 heavy atoms in consideration of the druggability~\cite{xiong2019pushing}.
Next, according to the concept of AC, we focus on the substructure modifications of only limited size and used chemically intuitive upper transformation size boundaries.
Specifically, the maximal size of modified substituents and the maximal difference between these substituents are limited to 13 and 8 non-hydrogen atoms, respectively.
The 8-atom restriction corresponds to the size of a substituted six-membered ring and the 13-atom restriction corresponds to the size of a substituted condensated two-ring systems~\cite{hu2012mmp}.

\clearpage
\section{More information of the ACNet.}
\label{sec:moreinformation}
\begin{figure}[htbp]
\begin{center}
\centerline{\includegraphics[width=12cm]{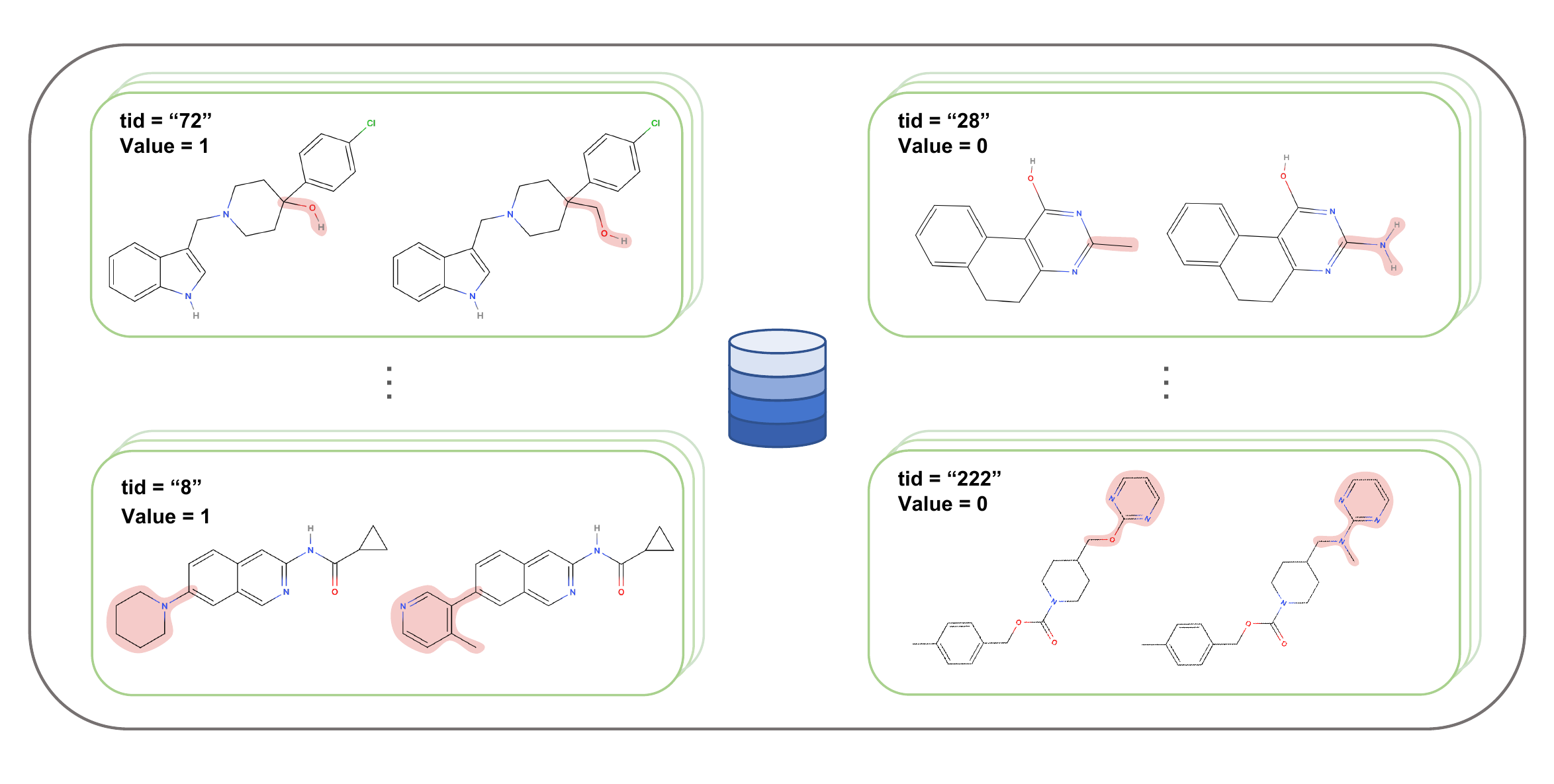}}
\caption{Examples of the data in the ACNet dataset. Data are organized by the target which the MMPs are against. Each sample contains the structures of two molecules (recorded by SMILES), the label and the target id.}
\label{fig:dataset}
\end{center}
\end{figure}

\begin{figure}[htbp]
\begin{center}
\centerline{\includegraphics[width = \textwidth]{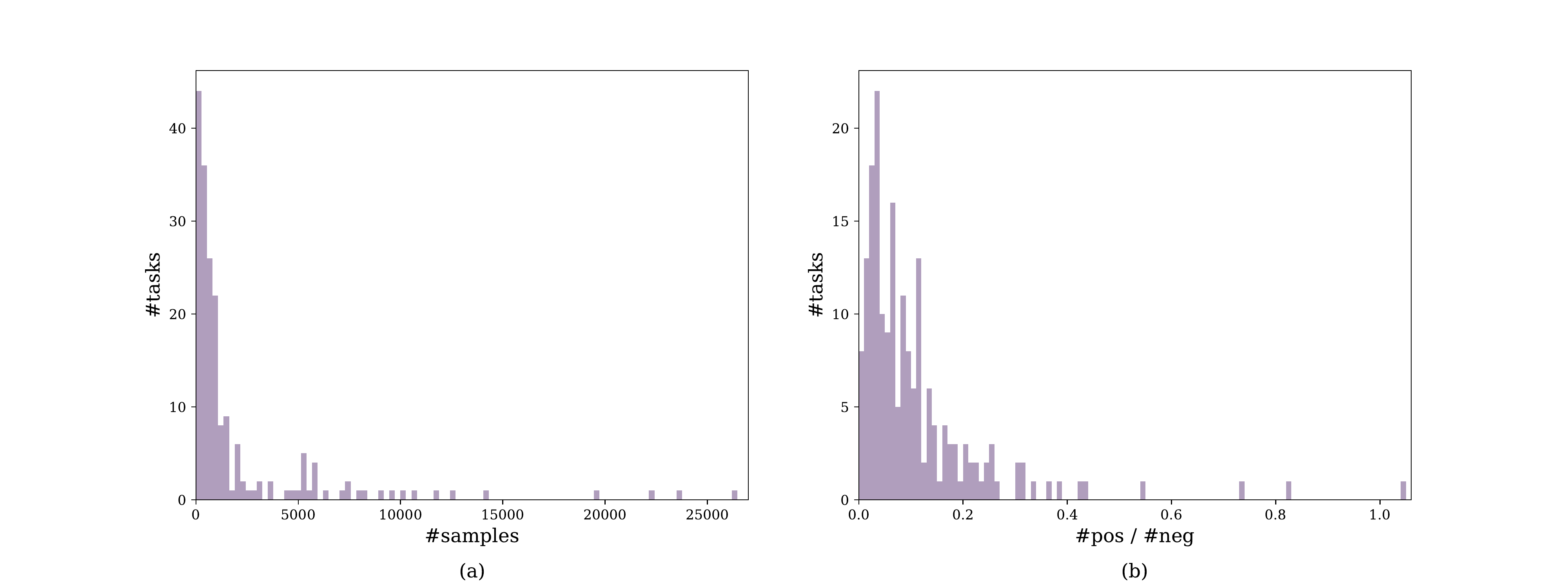}}
\caption{(a) The distribution of the number of samples of the tasks in the ACNet. (b) The distribution of the ratio $\frac{\#pos}{\#neg}$ of the tasks in the ACNet.}
\label{fig:distribution}
\end{center}
\end{figure}

Examples of the data in ACNet dataset are shown in Fig.~\ref{fig:dataset}.
Samples are organized by the target IDs, i.e. targets that MMPs are against.
Each sample consists of two molecules (recorded by SMILES), the labels and the target ID.
Fig.~\ref{fig:distribution}(a) presents the distribution of the number of samples of the tasks in the ACNet.
The distribution shows the low-data feature of the ACNet dataset.
And Fig.~\ref{fig:distribution}(b) presents the distribution of the ratio between the number of positive samples and the negative samples.
As shown in the figure, the ratio of most tasks is lower than 0.2, which indicates the imbalanced feature of the ACNet dataset.
Coupled with the OOD feature of the Mix subset, the tasks in ACNet are imbalanced, low-data and out-of-distribution, which is challenging for deep learning models.



\clearpage
\section{Information of baseline models}
Tab.~\ref{tab:models} shows the choice of models evaluated in the benchmark experiments.
These models are used as encoders in the baseline framework to extract molecular representations.
For each model, the subset experiments in which the model participates is ticked.
As the size of tasks in the Few subset is too small to train a deep learning model from scratch, self-supervised pre-trained models are involved as frozen molecular representation encoders.

\begin{table}[htbp]
  \caption{Choice of models to extract molecular representations.}
  \label{tab:models}
  \centering
  \begin{tabular}{ccccccc}
    \hline
    Category    &  Model  & Large & Medium & Small & Few & Mix \\
    \hline
    FingerPrints & ECFP+MLP~\cite{rogers2010extended} & \checkmark & \checkmark & \checkmark & \checkmark & \checkmark\\
    LM & LSTM~\cite{hochreiter1997long} & \checkmark & \checkmark & \checkmark & - & - \\
       & GRU~\cite{cho2014properties} & \checkmark & \checkmark & \checkmark & - & \checkmark \\
    GNNs & GCN~\cite{kipf2016semi} & \checkmark & \checkmark & \checkmark & - & \checkmark \\
         & GIN~\cite{xu2018powerful} & \checkmark & \checkmark & \checkmark & - & - \\
         & SGC~\cite{wu2019simplifying} & \checkmark & \checkmark & \checkmark & - & - \\
         & CMPNN~\cite{song2020communicative} & \checkmark & \checkmark & \checkmark & - & - \\
         & Graphormer~\cite{ying2021transformers} & \checkmark & \checkmark & \checkmark & - & \checkmark \\
    \hline
    Pre-trained Models & GROVER~\cite{rong2020self} & - & - & - & \checkmark & - \\
                       & ChemBERT~\cite{kim2021merged} & - & - & - & \checkmark & - \\
                       & MAT~\cite{maziarka2020molecule} & - & - & - & \checkmark & - \\
                       & PretrainGNNs~\cite{hu2019strategies} & - & - & - & \checkmark & - \\
                       & Pretrain8~\cite{chen2021extracting} & - & - & - & \checkmark & - \\
                       & S.T.~\cite{honda2019smiles} & - & - & - & \checkmark & - \\
                       & GraphLoG~\cite{xu2021self} & - & - & - & \checkmark & - \\
                       & GraphMVP~\cite{liu2021pre} & - & - & - & \checkmark & - \\
    \hline
  \end{tabular}
\end{table}

\clearpage

\section{Distribution of unique cores and substituents}
The distribution of unique cores and substituents and their and their occurrences in the dataset are computed.
Specifically, we counted the number of unique core and substituent substructures for each task, and computed the average of each subset.
The average number of samples (i.e. $avg( \# MMPs)$) are also computed.
For each task, the core diversity and substituent diversity are calculated by $\frac{\#unique~core}{\#MMPs}$ and $\frac{\#unique~ substituent}{\#MMPs}$, respectively.
Then, the average diversities over the entire subset are computed and reported.
These statistical measurements are shown in Tab.~\ref{tab:distribution_cs}.


\begin{table}[htbp]
  \caption{Distribution of unique cores and substituents in the subsets.}
  \label{tab:distribution_cs}
  \centering
  \resizebox{\linewidth}{!}{
  \begin{tabular}{ccccccc}
    \toprule
    Subsets & \#tasks & $avg(\#MMPs)$ & $avg(\#unique~core)$  &  $avg(\#unique~substituents)$ & $avg$ core diversity & $avg$ substituent diversity \\
    \midrule
    Large & 3 & 24078 & 976 & 816 & 0.0402 & 0.0339\\
    Medium & 64 & 4311 &459&397 & 0.1213 & 0.1227 \\
    Small & 110 & 483 & 68 & 106 & 0.1504 & 0.2511\\
    Few & 13 & 64 & 16 & 34 & 0.2592 & 0.5333\\
    
    \bottomrule
  \end{tabular}
  }
\end{table}

\clearpage
\section{Influence of core/substituent splitting}
\label{sec:cs_splitting}
In Sec.~\ref{sec:exp}, predictive performance of deep molecular representations have been benchmarked and evaluated.
As all of the related works introduced in Sec.~\ref{sec:related_work} have leveraged the core/substituent splitting method, in this section, we will conduct an supplementary experiment to show the influence of this splitting skill.
Specifically, given a pair of molecules in an MMP sample, their common core substructure and the two substituent substructures are computed.
Then, embeddings of these three substructures are encoded separately, and these embeddings are concatenated as input of an MLP for prediction.

GRU and LSTM models are used as backbone encoders.
Experiments are conducted on Small subset.
In addition, the previous work by Iqbal \textit{et al.}~\cite{iqbal2021prediction} are implemented and tested on Small subset, either.
In their work, three substructures are transformed into the three $300\times 300$ \textbf{images} of graphs by RDKit package, and a CNN is used for encoding the concatenated $300 \times 900$ image.
As their source codes are not publicly available, we have implemented their model with structural hyperparameters introduced in their article, and searched for other hyperparameters.
The results are shown in Tab.~\ref{tab:separate}, where results of models without splitting are cited from Tab.~\ref{tab:baseline_exp} for comparison.
Models without splitting are denoted by \textit{no sp}, and models with splitting are denoted by \textit{sp}.

The experimental results from Tab.~\ref{tab:separate} reveal a significant improvement of predictive accuracy for GRU and LSTM models.
With splitting method, over 6 percentage of improvement can be achieved by GRU and LSTM.
And Iqbal~\textit{et al.}'s model shows better performance than other deep learning models reported in Sec.~\ref{sec:exp}.
These results demonstrate that the splitting method can serve as an add-on skill to improve prediction accuracy in practice.
In addition, these models with splitting method involved still cannot outperform the ECFP+MLP model.
It further demonstrates that the ECFP is more appropriate as molecular representations for the MMP-cliff prediction task.

\begin{table}[htbp]
  \caption{Experimental results of models with core/substituent splitting method on Small subset.}
  \label{tab:separate}
  \centering
  \begin{tabular}{cc|cc|c}
    \toprule
    \multicolumn{2}{c|}{GRU} & \multicolumn{2}{c|}{LSTM} & Iqbal~\textit{et al.}'s model             \\
    \midrule
    no sp & sp & no sp & sp & sp\\
    0.803 $\pm$ 0.007 & 0.863 $\pm$ 0.016 & 0.812 $\pm$ 0.008 & 0.878 $\pm$ 0.005 & 0.861 $\pm$ 0.003\\
    \bottomrule
  \end{tabular}
\end{table}

\end{document}